# Vortex trapping and expulsion in thin-film type-II superconducting strips

K. H. Kuit, J. R. Kirtley, J. R. Clem, H. Rogalla, J. Flokstra

*Abstract*—Vortex trapping is investigated in thin-film strips of superconducting material. We present a model for the critical field above which vortex trapping occurs in these strips. This model includes the pairing energy of vortex-antivortex pairs in addition to the energy of single vortices. Experimental verification of the model with a scanning SQUID microscope shows very good agreement between the model and experiments on $YBa_2Cu_3O_{7-\delta}$ and Nb strips. Statistical analysis of the vortex distribution in the strips above the critical field has been performed and a comparison has been made between Nb and $YBa_2Cu_3O_{7-\delta}$ for the distributions in the lateral and longitudinal directions.

## I. Introduction

PINNED or trapped vortices are nearly always observed in thin-film type-II superconductors, even when cooled in relatively low magnetic fields. In general, this can be attributed both to pinning of vortices by, for example, defects and grain boundaries, and to trapping by geometric energy barriers. Understanding such pinning and trapping effects is important for superconducting electronics applications.

Flux trapping plays an important role in the properties of superconducting magnetic field sensors like high-$T_c$ SQUIDs [1,2] and hybrid magnetometers based on a high-$T_c$ flux concentrator [3,4,5]. When these devices are operated in an unshielded environment such as, for example, in biomagnetism, geophysical research, or space applications, the low-frequency sensitivity of the sensor is limited by thermal hopping of trapped vortices in the superconducting body, which gives rise to 1/f noise. This noise can be eliminated by dividing the high-$T_c$ body into thin strips [1,2,5]. Research on SQUIDs with slotted or meshed washers show that indeed the low-frequency sensitivity can be improved. The resulting superconducting strips have a certain critical induction below which no vortex trapping occurs, resulting in an extended ambient field range in which these sensors can be effectively operated. We investigated vortex trapping in thin-film $YBa_2Cu_3O_{7-\delta}$ (YBCO) strips in order to incorporate the results in a hybrid magnetometer based on a YBCO ring tightly coupled to, for example, a giant-magneto-resistance (GMR) or Hall sensor. We compare the results on YBCO strips with our similar new measurements on Nb strips in order to study the influence of material properties like coherence length and growth morphology.

## II. Theory

Whether or not a vortex gets trapped in an infinitely long superconducting thin-film strip of width $W$ cooled in a background magnetic field $B_a$ is determined by the Gibbs free energy [6]:

$$G(x) = \frac{\Phi_0^2}{2\pi\mu_0\Lambda}\ln\left[\frac{\alpha W}{\xi}\sin\left(\frac{\pi x}{W}\right)\right] \mp \frac{\Phi_0(B_a - n\Phi_0)}{\mu_0\Lambda}x(W-x). \quad (1)$$

Here $\Phi_0$ is the flux quantum, $\mu_0$ the permeability of vacuum, $\xi$ the coherence length, n the vortex density, and $x$ the lateral position in the strip with $0 \le x \le W$. $\Lambda$ is the Pearl length given by $\Lambda = 2\lambda^2/d$, where $\lambda$ is the London penetration depth and d the thickness of the film. $\alpha$ is a constant factor determined by an assumption about the core size of the vortex. The Gibbs free energy consists of two terms. The first term is the self-energy term and has a dome shape. The second term, slightly modified as explained in [7], is the field interaction term with a vortex (upper sign) or an antivortex (lower sign) and has a parabolic shape. Dependent on the applied magnetic background field during cooling, a dip can occur in the Gibbs free energy, which can act as an energy barrier for the escape of vortices. The normalized Gibbs free energy is displayed for a number of magnetic fields in Fig 1.

Prior to the work of [7] there were two existing models for the critical field above which vortex trapping will occur, both of which use (1) in the limit of n→0. The first model, presented in [6], assumes a metastable condition, i.e., that trapping first occurs when there is a dip in the Gibbs free energy. This condition occurs when $d^2G(W/2)/dx^2 = 0$, resulting in the relation:

Manuscript received 15 August 2008. This work was funded by the Dutch MicroNED program. The scanning SQUID microscope used in this research was donated to the University of Twente by IBM T. J. Watson Research Center.

K. H. Kuit, H. Rogalla and J. Flokstra are with the Low Temperature Division, Mesa+ Institute for Nanotechnology, University of Twente, Enschede, The Netherlands. (corresponding author K. H. Kuit phone: +31 53-4894627; e-mail: k.h.kuit@tnw.utwente.nl).
J. R. Kirtley is with the Department of Applied Physics, Stanford University, Palo Alto, California 94305, USA, supported by the Center for Probing the Nanoscale, a NSF NSEC, NSF Grant No. PHY-0425897.
J. R. Clem is with the Ames Laboratory–DOE and Department of Physics and Astronomy, Iowa State University, Ames, Iowa 50011, USA, supported by the Department of Energy - Basic Energy Sciences under Contract No. DE-AC02-07CH11358.

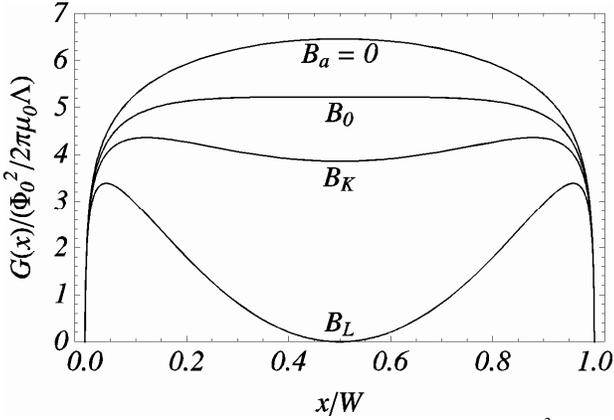

Fig. 1. Gibbs free energy of an isolated vortex in units of $\Phi_0^2/2\pi\mu_0\Lambda$ versus the strip width for applied inductions of $B_a=0$, $B_0=\pi\Phi_0/4W^2$, $B_K=1.65\Phi_0/W^2$ and $B_L=(2\Phi_0/\pi W^2)\ln(2W/\pi\xi)$ for n=0 and $\xi/W=10^{-3}$.

$$B_0 = \frac{\pi\Phi_0}{4W^2}. \quad (2)$$

In this case $\alpha=2/\pi$ [6] is assumed.

The second model, presented in [8], considers an absolutely stable condition. This happens when the energy in the middle of the strip equals zero and results in:

$$B_L = \frac{2\Phi_0}{\pi W^2}\ln\left(\frac{\alpha W}{\xi}\right). \quad (3)$$

Recently [7] we presented a new model, which involves the energy required to form vortex-antivortex pairs. Just below the critical temperature $T_c$ thermal fluctuations cause the generation of vortex-antivortex pairs. It follows from (1) that the Gibbs free energy for an antivortex does not exhibit a dip, and so antivortices can easily escape from the strip. However, thermally generated vortices have to overcome a magnetic-field-dependent energy barrier in order to escape. The rate of generation of vortex-antivortex pairs depends upon the pairing energy [9]:

$$E_{pair} = \frac{\Phi_0^2}{4\pi\mu_0\Lambda}. \quad (4)$$

In our model we assume that at the critical field the height of the energy barrier should equal the pairing energy of the vortex-antivortex pair, which gives the relation:

$$\max\left\{\ln\left[\sin\left(\frac{\pi x}{W}\right)\right] + \frac{2\pi B_a}{\Phi_0}\left[\frac{W^2}{4} - x(W-x)\right]\right\} = \frac{1}{2}, \quad (5)$$

where the maximum on the left-hand side is taken with respect to $x$. This equation can be solved numerically, resulting in a critical field [7]:

$$B_K = 1.65\frac{\Phi_0}{W^2}. \quad (6)$$

Also in this case $\alpha=2/\pi$ is assumed. Note that the critical fields presented in (2) and (6) are dependent only on the width of the strip but that the critical field in (3) also contains the coherence length. In Fig. 1 the Gibbs free energy as a function of $x$ is displayed for zero field and for the critical fields of (2), (3) and (6) with $\xi/W=10^{-3}$.

So far the Gibbs free energy was used in the limit of $n\to0$. However it is possible to derive a relation from (1) for the vortex density as a function of the applied magnetic field [7]:

$$n = \frac{B_a - B_K}{\Phi_0}. \quad (7)$$

A more thorough derivation of the equations presented in this section can be found in [7].

III. EXPERIMENTS

Measurements with a scanning SQUID microscope (SSM) [10] have been performed on YBa$_2$Cu$_3$O$_{7-\delta}$ (YBCO) and Nb strips. The SQUID used in the SSM had a pickup loop that was defined by focused ion beam milling and had an effective area of 10-15 μm$^2$. A solenoid coil around the SQUID and sample was used to apply a magnetic field to the sample. A triple mu-metal shield is present around the setup to eliminate the earth's magnetic field. The small residual magnetic field is compensated by the solenoid coil.

The YBCO sample was prepared by a pulsed laser deposited layer of YBCO on a substrate of SrTiO$_3$. The sample was structured by Ar ion etching. The Nb sample was made with a dc-sputtered Nb film on SrTiO$_3$, which was also structured with Ar ion etching. For both types of samples the deposited layer was approximately 200 nm thick, and the etching was performed at a 45° angle along the length of the strip for a relatively high etching rate.

The samples contain strips in varying width from 2-50 μm and have been cooled down in a large number of magnetic fields. The actual vortex trapping takes place just below $T_c$, but for the measurement the sample is further cooled to $T$=4.2K, the operating temperature of the SQUID. This further cooling does not affect the vortex trapping. In between cooling cycles the sample was warmed up to well above $T_c$.

A. Critical magnetic field vs. strip width

The results of the measurement of the critical field vs. strip width are displayed in Fig. 2 together with the critical field values of (2), (3) and (6). In the case of the model presented in (3) an estimation has to be made for the temperature-dependent coherence length. In this case one needs to know the trappinng temperature $T_{tr}$, the temperature at which the vortices are pinned in their final positions. Trapping temperatures of $T/T_c$=0.98 [11] and $T/T_c$=0.9985 [12] are used for YBCO and Nb respectively. YBCO has a coherence length of $\xi_{YBCO}(T$=0K$)$ = 3 nm, and the coherence length at the trapping temperature can be calculated as $\xi_{YBCO}(T_{tr})$ = 10.39 nm using the two-fluid model. The coherence length of Nb is $\xi_{Nb}(T$=0K$)$ = 38.9 nm, which results in $\xi_{Nb}(T_{tr})$ = 320 nm. Two curves are displayed in Fig. 2 for the critical field in (3) using $\alpha= 2/\pi$ [6] and $\alpha=1/4$ [8]. There are two data points in Fig. 2 for each strip width. The upper points indicate the lowest possible magnetic field where there are still some vortices present in the strip. The lower points show the highest possible field without any vortices visible. These represent upper and lower bounds for the actual critical field.

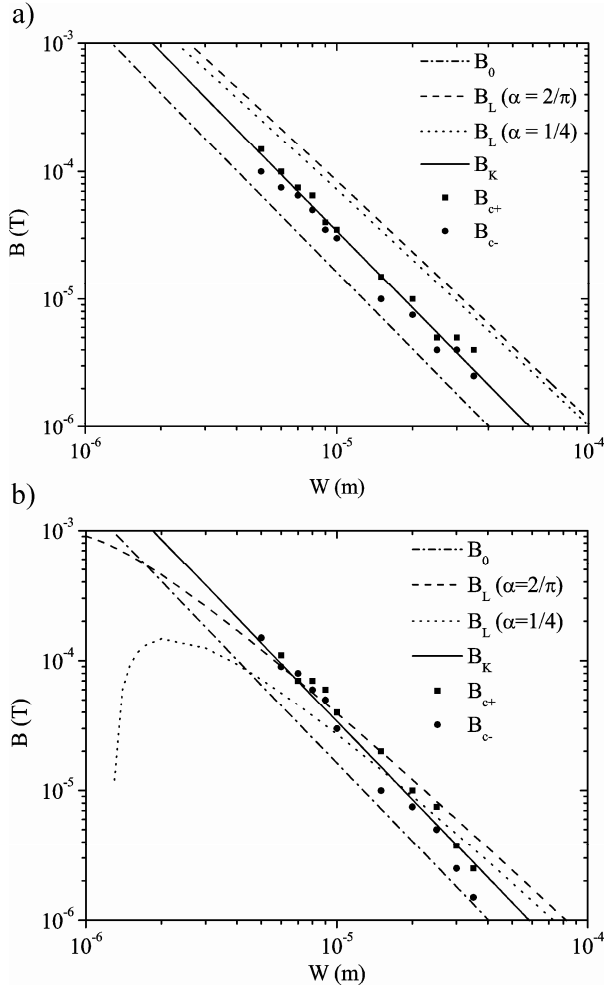

Fig. 2. Critical field versus strip width for a) YBCO and b) Nb. The squares $B_{c+}$ represent the lowest field at which vortex trapping occurs and the dots $B_{c-}$ the highest field without visible vortex trapping. The dot-dashed line shows $B_0$ (2), the solid line $B_K$ (6), and the dashed and dotted lines represent $B_L$ (3) for two different values of α.

For the measurements on YBCO in Fig. 2a) there is good agreement between the measurements and the critical field values of our model in (6). The values of both the metastable and absolutely stable conditions do not correspond to the measurement. In Fig. 2b) the field values of the model of Likharev in (3) are closer to our model. Even though there are solutions for the field values of (3) for both core size assumptions α, the field values of our model show a much better fit over the whole width range. It can be concluded from the measurements on YBCO and Nb that the critical field depends only on the width of the strip and not on the coherence length.

### B. Vortex ordering

Vortices are trapped in the strips for fields above the critical field. The minimum in the Gibbs free energy in the center of the strip makes it energetically favorable for the vortices to be trapped there. As the magnetic field is increased, more vortices get trapped in the strip according to (7) and the vortex-vortex interaction becomes dominant, resulting in trapping in two parallel rows. Simulations on this topic are presented in [13], where a transition field from one to two rows is given to be $B_a=2.48B_c$. The transition field is verified for YBCO strips and is presented in [7].

In this paper we compare the trapping distribution in the single-row regime for YBCO and Nb. Typical SSM images can be seen on YBCO in Fig. 3a) and Nb in Fig. 3b). This particular case shows 30 μm wide strips cooled down in approximately 10 μT (the field for the Nb strips was slightly higher).

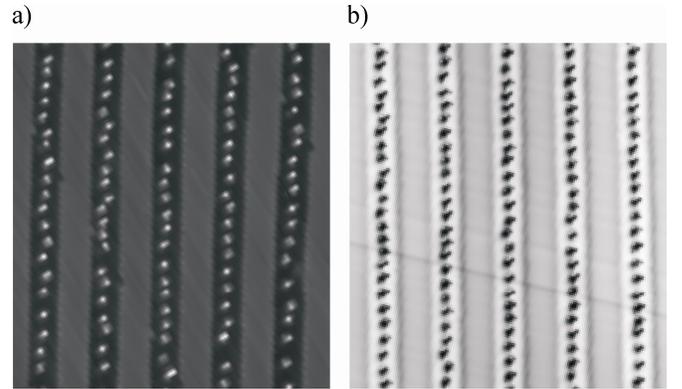

Fig. 3. SSM images of 30 μm wide strips of a) in YBCO and b) in Nb cooled down in a magnetic field of approximately 10 μT. The field for the Nb strips was slightly higher than for the YBCO strips.

From these images it is evident that the ordering of the vortices differs for the two materials. The vortices are more homogeneously distributed in the center of the strips for the Nb strips. Also the spacing along the length of the strip is more homogeneous compared to YBCO.

Statistical analyses were carried out in lateral and longitudinal direction to compare the different trapping distributions.

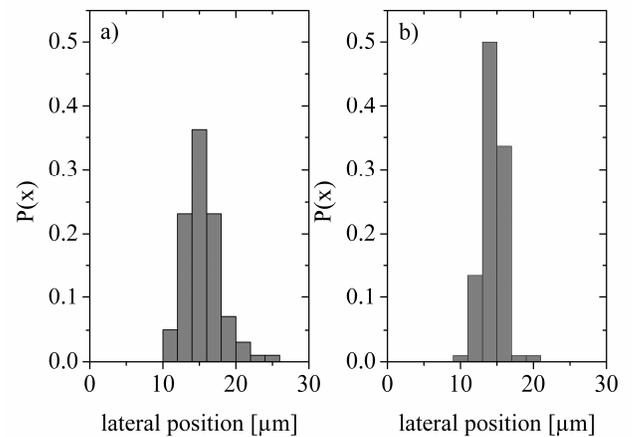

Fig. 4. Histograms of the lateral vortex distribution of 30 μm wide strips at 10 μT for a) YBCO and b) Nb. The histogram for the Nb strips is evidently narrower with a standard deviation of 1.42 μm compared to 2.58 μm for the YBCO strips.

In Fig. 4 the histograms of the lateral distribution for YBCO in Fig. 4a) and Nb in Fig. 4b) are displayed. Obviously the trapping positions are more narrowly distributed around the center of the strip in Nb. The standard deviation (STD) of the histograms is used to provide a quantitative measure of this distribution. For YBCO and Nb the STDs are respectively 2.58 μm and 1.42 μm. Comparison of the STDs at different magnetic fields showed no field dependency and were always of comparable values.

The longitudinal vortex distribution for the two materials is investigated as a function of the magnetic field. In Fig. 5 the normalized standard deviation versus the vortex density $n$ is displayed.

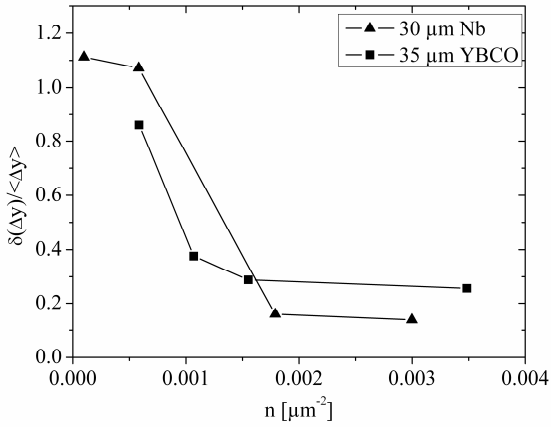

Fig. 5. Normalized standard deviation of the longitudinal vortex distribution versus the width for 30 μm Nb strips and 35 μm YBCO strips. The normalized standard deviation is the standard deviation divided by the mean distance between the vortices.

The dataset used in this analysis did not have exactly the same strip width for the two materials. However, when the longitudinal vortex distribution is analyzed versus the vortex density, this is not important. For both materials the longitudinal ordering is improving when the field is increased, which means that the STD is decreasing faster than the average distance between the vortices. For relatively high fields the ordering is better for Nb than for YBCO. For relatively low fields there is not enough data available to draw strong conclusions.

As was mentioned previously, the YBCO layer was deposited by laser ablation at a substrate temperature of 780°C, resulting in a polycrystalline film with domains of ~200 nm. The Nb film was sputtered at room temperature, also resulting in a polycrystalline film but with smaller domains of ~20 nm. We believe that the combination of smaller grains and larger coherence length leads to larger coupling between the domains in Nb than in YBCO. We therefore do not expect a large number of deep additional wells in the Gibbs free energy for Nb. On the other hand, YBCO is a more complex material, which will exhibit more defects and we assume many more deep additional wells. This could explain the larger STDs in the YBCO strips in the lateral as well as the longitudinal direction.

## IV. Conclusions

Measurements with an SSM on the critical field for vortex trapping in YBCO and Nb show that our model, which includes the vortex-antivortex pairing energy, gives the best fit. Furthermore it can be concluded from these measurements that the critical field is dependent only on the width of the strip and not on the material. Statistical analysis on the vortex distribution shows that the trapping in Nb is much more ordered than for YBCO in both the lateral and longitudinal directions. We believe that YBCO has many more severe defects leading to deep additional energy wells in the Gibbs free energy. These strong pinning sites have a negative influence on the ordering of the vortices.